\documentclass[cits]{PoS}
\newcommand{\beq}{\begin{equation}}
\newcommand{\eeq}{\end{equation}}
\newcommand{\bea}{\begin{eqnarray}}
\newcommand{\eea}{\end{eqnarray}}

\newcommand{\ba}{\begin{array}}
\newcommand{\ea}{\end{array}}

\newcommand{\req}[1]{eq.~(\ref{#1})}

\newcommand{\rep}[1]{\cite{#1}}

\newcommand{\dif}{{\rm d}}

\newcommand{\Dslash}{\relax{\kern+.25em / \kern-.70em D}}

\newcommand{\Real}{\relax{\mathsf{\Gamma\kern-.35em R}}}
\newcommand{\Int}{\relax{\mathsf{Z\kern-.40em Z}}}

\newcommand{\gbar}{\kern1pt\overline{\kern-1pt g\kern-0pt}\kern1pt}

\newcommand{\mbar}{\kern2pt\overline{\kern-1pt m\kern-1pt}\kern1pt}

\newcommand{\obar}[1]{\kern3pt\overline{\kern-2pt #1\kern-0pt}\kern1pt}

\newcommand{\corrbar}[1]{\kern3pt\overline{\kern-2pt #1\kern-0pt}\kern1pt}

\newcommand{\oVApAV}[1]{#1_{\rm\scriptscriptstyle VA+AV}}

\newcommand{\oVVpAA}[1]{#1_{\rm\scriptscriptstyle VV+AA}}

\newcommand{\oVApAVren}[1]{\kern3pt\overline{\kern-2pt #1\kern-0pt}\kern1pt_{\rm\scriptscriptstyle VA+AV;s}}

\newcommand{\ZVApAV}[1]{Z_{\rm\scriptscriptstyle VA+AV #1}}

\newcommand{\ZtotVApAV}[1]{\mathcal{Z}_{\rm\scriptscriptstyle VA+AV #1}}

\newcommand{\zbar}{\kern3pt\overline{\kern-2pt Z\kern-0pt}\kern1pt}

\newcommand{\zbarVApAV}[1]{\kern3pt\overline{\kern-2pt Z\kern-0pt}\kern1pt_{\rm\scriptscriptstyle VA+AV #1}}
\newcommand{\zrgiVApAV}[1]{\hat Z_{\rm\scriptscriptstyle VA+AV #1}}

\newcommand{\sigVApAV}[1]{\sigma_{\rm\scriptscriptstyle VA+AV #1}}

\newcommand{\SigVApAV}[1]{\Sigma_{\rm\scriptscriptstyle VA+AV #1}}

\newcommand{\UVApAV}[1]{U_{\rm\scriptscriptstyle VA+AV #1}}

\newcommand{\mumin}{\mu_{\rm min}}

\newcommand{\cF}{{\cal F}}

\newcommand{\cO}{{\cal O}}

\title{Non-perturbative scale evolution of four-fermion operators in two-flavour QCD
\thanks{Preprint: ROM2F/2006/23, MKPH-T-06-09, CERN-PH-TH/2006-202, TRINLAT-06/05}
}

\ShortTitle{Non-perturbative scale evolution of four-fermion operators in two-flavour QCD}

\author{\includegraphics[width=.15\textwidth]{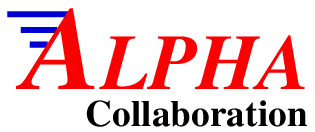}}

\author{Petros Dimopoulos, \speaker{Gregorio Herdoiza}, Anastassios Vladikas\\
INFN, Sezione di Roma Tor Vergata\\
    and Dipartimento di Fisica, Universit\`a di Roma ``Tor Vergata''\\
    Via della Ricerca Scientifica 1, I-00133 Rome, Italy\\
    E-mail: \email{\{dimopoulos,herdoiza,vladikas\}@roma2.infn.it}
    }

\author{Filippo~Palombi\\
Institut f\"ur Kernphysik,
    Johannes Gutenberg Universit\"at\\
    Johann Joachim Becher-Weg 45,
    D-55099 Mainz, Germany\\
    E-mail: \email{palombi@kph.uni-mainz.de}
}

\author{Carlos~Pena\\
CERN, Physics Department, TH Division,
    CH-1211 Geneva 23, Switzerland\\
    E-mail: \email{carlos.pena.ruano@cern.ch}
}

\author{Stefan~Sint\\
School of Mathematics,
    Trinity College,
    Dublin 2,
    Ireland \\
    E-mail: \email{sint@maths.tcd.ie}
}

\abstract{
We apply finite-size recursion techniques
based on the Schr\"odinger functional formalism
to determine the renormalization group running of
four-fermion operators which appear in the 
$\Delta S=2$ effective weak Hamiltonian
 of the Standard Model. Our calculations are
done using  $\cO(a)$ improved Wilson fermions with 
$N_{\rm f}=2$ dynamical flavours. 
Preliminary results are presented for the four-fermion operator
 which determines the $B_K$-parameter in
tmQCD. 
}

\FullConference{XXIVth International Symposium on Lattice Field Theory\\
                July 23-28, 2006\\
                Tucson, Arizona, USA}

\begin{document}

\section{Introduction}

A precise determination of the Kaon $B$-parameter is required to constrain
 the CKM unitarity triangle analysis.
$B_K$ is defined in terms of  hadronic matrix elements which
 can be computed using lattice QCD. 
At present, the inclusion of  dynamical quark effects is an essential requirement
 in these lattice calculations.
Indeed, a remarkably good agreement has been found between several independent
 quenched determinations of
$B_K$~\cite{Dawson:2005za}, leaving the ``quenching'' effects as the 
largest uncertainty. A complete
study of the systematic effects, other than quenching, has been performed in 
ref.~\rep{Dimopoulos:2006dm} and
reviewed in this conference~\cite{Carlos:Lat06}. 
There are indications
 that once dynamical quarks 
are taken into account, a major source of  uncertainty on lattice results arises 
from the determination of the  renormalization
factors of the four-fermion operators~\cite{Dawson:2005za}. 
In principle, this uncertainty can be completely eliminated by a
non-perturbative renormalization procedure. Similarly,
the renormalization group (RG) running of the operator from hadronic scales up to
high-energies, where perturbation theory can be safely applied,
is best performed non-perturbatively.
Here we report on the status of such a non-perturbative 
renormalization using $N_{\rm f}=2$ dynamical 
flavours. In particular, we present 
preliminary results for the scale
evolution of
the four-fermion operator relevant for the determination of $B_K$ 
in the context of
tmQCD~\cite{Frezzotti:2000nk}.

The theoretical description of the $K^0 - \bar K^0$ oscillation is 
controlled, once high-energy
scales are integrated through an operator product expansion 
procedure, by the matrix element 
$\langle \bar K^0 \vert O^{\Delta S = 2} \vert K^0 \rangle$, where 
the four-fermion operator is defined as
follows:
%%%
\begin{eqnarray}
O^{\Delta S = 2} &\equiv& [\bar s \gamma_\mu (1-\gamma_5) d] \,\,
[\bar s \gamma_\mu (1-\gamma_5) d]\,\, = \,\, \oVVpAA{O} - \oVApAV{O} \, .
\end{eqnarray} 
%%%
The strange and down quark fields are denoted by $s$ and $d$ respectively.
The $B_K$-parameter is expressed in terms of the parity-even
 $\oVVpAA{O}$ operator~:
\begin{eqnarray}
B_K \equiv  \frac{\langle \bar K^0 \vert \oVVpAA{O} \vert K^0 \rangle}
{\frac{8}{3} F_K^2 m_K^2} \, .
\end{eqnarray}
Parity conservation ensures that the matrix element 
$\langle \bar K^0 \vert \oVApAV{O} \vert K^0 \rangle$,
involving  the parity-odd operator, 
is identically zero.

In lattice regularizations preserving chiral symmetry the 
operator $\oVVpAA{O}$ is multiplicatively
renormalizable. This is not the case for  Wilson fermions because in this case chiral symmetry is broken
at non-zero lattice spacing. The renormalization of $\oVVpAA{O}$ is 
therefore more involved since mixing with four other dimension-6 operators has to be
 considered. On the other hand, discrete
symmetries protect the parity-odd operator $\oVApAV{O}$, so as to 
preserve multiplicative
renormalization also in the case of Wilson 
fermions~\cite{Bernard:1987pr,Donini:1999sf}.

The inclusion of a ``twisted mass'' term in the fermionic action 
opens the way to improve the renormalization
properties of Wilson fermions. The twisted-mass theory is related 
to  standard QCD through an axial
transformation of the quarks fields. By choosing appropriate
formulations of tmQCD
(see ref.~\cite{Dimopoulos:2006dm} for two of these formulations), it is
possible to relate the  QCD $\oVVpAA{O}$ operator  to a partner
$\oVApAV{O}$ in 
tmQCD where the property of
multiplicative renormalization of this operator is preserved. This
method allows the determination of $B_K$ only
via multiplicative renormalization, thus avoiding mixing with operators of
wrong chirality. Another proposal in this direction uses chiral 
Ward-Takahashi identities to relate parity-odd
and parity-even operators~\cite{Becirevic:2000cy}.

Our strategy to renormalize the  operator $\oVApAV{O}$ closely
follows the one used in ref.~\rep{Guagnelli:2005zc} for the 
quenched case. The connection between the
renormalization group invariant operator $\hat \oVApAV{O}$ and its
bare counterpart $\oVApAV{O}(g_0)$ can be written 
in the following way:
%%%
\begin{equation}
 \hat \oVApAV{O}(x) =
\lim_{g_0 \rightarrow 0} \,\,\, \ZtotVApAV{;s}(g_0) 
\,\,\,\oVApAV{O}(x;g_0)\,.
\label{eq:Ototren}
\end{equation}
%%%
The RGI operator is independent of the renormalization scheme 
and scale when the renormalization conditions are imposed at zero
quark mass~\cite{Weinberg1973}. The renormalization factor
$\ZtotVApAV{;s} (g_0)$ is scale-independent but depends on the 
scheme $s$ (only through cutoff effects) and on
the lattice regularization. These dependences are manifest 
when decomposing $\ZtotVApAV{;s} (g_0)$ into:
%%%
\begin{equation}
  \label{eq:tot_renorm}
  \ZtotVApAV{;s}(g_0) =\zrgiVApAV{;s}(\mumin)\,
  \ZVApAV{;s}(g_0,a\mumin) \ .
\end{equation}
%%%
The first factor on the r.h.s.~of \req{eq:tot_renorm} controls
 the RG-running of the operator from the
reference scale $\mumin$ to an infinite scale. It is independent of the regularisation.
 The second factor, $\ZVApAV{;s}(g_0,a\mumin)$, 
relates the bare lattice operator to its continuum
value at the hadronic scale $\mumin$. This factor is therefore 
dependent on both the regularization and the
scale. 
Both factors on the r.h.s.~of \req{eq:tot_renorm} depend on the renormalization scheme.
In this report, we focus on the more  expensive part of 
the renormalization procedure, that is, the
computation of the contribution of the non-perturbative evolution
 function $\zrgiVApAV{;s}(\mumin)$ 
describing the running in the scale range
 $1$~GeV~--~$100$~GeV. 

\section{Renormalization group running of four-fermion operators}

Let us first define a renormalized four-fermion operator $O_R$ at
a reference scale $\mu$~:
%%%
\begin{equation}
 O_R(x; \mu) =
\lim_{a \rightarrow 0} \,\,\,  Z_{O}(g_0,a\mu) \,\,\,O(x;g_0)\,.
\label{eq:Oren}
\end{equation}
%%%
The running of the renormalized  operator $O_R(\mu)$ is controlled 
by its anomalous
dimension $\gamma_O(\bar g)$, defined as~:
%%%
\begin{equation}
\mu \,\,\, \frac{\partial}{\partial \mu} O_R(x;\mu) =
\gamma_O(\bar g) \,\,\, O_R(x;\mu) \ .
\label{eq:rg}
\end{equation}
%%%
In mass-independent renormalization schemes~\cite{Weinberg1973} as those we consider here, 
the function $\gamma_O(\bar g)$ only depends 
on the renormalized coupling $\bar g$. The perturbative expansion 
of $\gamma_O(\bar g)$ is given by
%%%
\begin{equation}
\gamma_O(g) \stackrel{g \to 0}{\sim}
-g^2\left(\gamma_O^{(0)} + \gamma_O^{(1)} g^2
+ \gamma_O^{(2)} g^4 + \ldots \right) \ ,
\end{equation}
%%%
with $\gamma_O^{(0)}$ a universal coefficient. 
By combining eqs.~(\ref{eq:Oren}) and (\ref{eq:rg}), it is
possible to relate
the anomalous dimension of $O_R$ to its scale-dependent 
renormalization factor~:
%%% 
\begin{equation}
\label{gamma_to_Z}
\gamma_O(\gbar(\mu)) = \lim_{a \to 0} 
\left(\mu \frac{\partial}{\partial \mu} Z_O(g_0,a\mu)\right)
Z_O(g_0,a\mu)^{-1} \ .
\end{equation}
%%%
The RGI
operator is obtained upon the formal
integration of eq.~(\ref{eq:rg}). It is given by
\begin{equation}
\hat O(x) \,\, = \,\, O_R(x;\mu) \,\,\, 
\left[\frac{\bar g^2(\mu)}{4\pi}\right]^{-\gamma_O^{(0)}/(2b_0)} 
\exp\left\{ - \int_0^{\bar g(\mu)} dg \left( 
\frac{\gamma_O(g)}{\beta(g)} - \frac{\gamma_O^{(0)}}{b_0 g} \right) \right\} \, .
\label{eq:rgi}
\end{equation}
%%%
The integral in the r.h.s~of \req{eq:rgi} describes the 
scale evolution of $\gamma_O(\bar g)$. The evolution
function of the operator $O_R$  between the renormalization
 scale $\mu$ and an arbitrary scale $\mu'$ is given by:
%%%
\begin{equation}
\label{RG_evolution}
U(\mu',\mu) \equiv \exp\left\{\int_{\gbar (\mu)}^{\gbar (\mu')}
\frac{\gamma_{O}(g)}{\beta(g)} \dif g
\right\} 
= \lim_{a \to 0} \frac{Z_{O}(g_0,a\mu^\prime)}{Z_{O}(g_0,a\mu)} 
\ .
\end{equation}
%%%
The running of the renormalized four-fermion operator $O_R$ 
can therefore  be performed by constructing ratios of
the renormalization factors $Z_{O}$ at different scales.

Our renormalization schemes are defined in the
 Schr\"odinger functional (SF) formalism. This technique 
 has been
used to determine the scale evolution of physical
 quantities  such as the strong
coupling~\cite{L\"uscher:1993gh,DellaMorte:2004bc} or the
 quark mass~\cite{Capitani:1998mq,DellaMorte:2005kg}. These
studies were performed both with and without dynamical
 quarks. In the case of four-fermion
operators~\cite{Guagnelli:2005zc} the running was 
carried out in the quenched approximation. We regularize the theory on a lattice 
of physical size $L^4$ using standard 
SF boundary conditions 
allowing to carry out simulations at zero quark masses. The SF is used 
as a mass-independent renormalization scheme; since the renormalization factors are  
flavour-independent, 
they can also be used to renormalize the $B$-parameters in 
the Kaon, $D$ and $B$-meson sectors. The renormalization 
conditions are imposed at a scale $\mu$ equal to the IR cutoff $1/L$.

Let us now concentrate on the case of the local parity-odd four-fermion operator:
%%%
\begin{equation}
  \oVApAV{O}(x)=\frac{1}{2}\left[
      (\bar{\psi}_{1}\gamma_\mu\psi_{2})
        (\bar{\psi}_{3}\gamma_\mu\gamma_5\psi_{4})
   +(\bar{\psi}_{1}\gamma_\mu\gamma_5\psi_{2})(\bar{\psi}_{3}\gamma_\mu\psi_{4})
   + (\psi_2 \leftrightarrow \psi_4) \right]\, .
 \label{eq:operators}
\end{equation}
%%%
Four distinct valence flavours are used in the definition of the operator.
The SF correlation functions used to extract the operator
 $\oVApAV{O}(x)$ are:
\begin{equation}
\cF_{[\Gamma_{\rm A},\Gamma_{\rm B},\Gamma_{\rm C}]} (x_0) = \frac{1}{L^3}
\langle \cO_{21}[\Gamma_{\rm A}]  \cO_{45}[\Gamma_{\rm B}]
\,\,\,  \oVApAV{O}(x) \,\,\,
\cO^\prime_{53}[\Gamma_{\rm C}] \rangle \, ,
\label{eq:f4-corr}
\end{equation}
where $\cO$ and $\cO^\prime$ are the interpolating fields on the time boundaries 
(refer to~\cite{Guagnelli:2005zc} for a full explanation of the 
 notations). Several choices of the Dirac matrices 
$\Gamma_{\rm A,B,C}$ are allowed. We will  focus here on the
particular choice $\Gamma_{\rm A}=\Gamma_{\rm B}=\Gamma_{\rm C}=\gamma_5$. 
Note that a ``spectator'' valence 
quark $\psi_5$ is used
in \req{eq:f4-corr}. It is useful to keep in mind that since
 the quarks are massless in this mass-independent
 renormalization scheme, flavour
 only enters through Wick contractions.
 
The logarithmic divergences of the local operator $\oVApAV{O}(x)$ are isolated by dividing out 
from the correlator $\cF$
the divergences coming from the boundaries and the external legs.
 This is obtained through the ratio:
%%%
\begin{equation}
h(x_0) = \frac{\cF_{[\gamma_5,\gamma_5,\gamma_5]}(x_0)}{f_1^{3/2}} ,
\label{eq:h-corr}
\end{equation}
%%%
where $f_1$ is the boundary-to-boundary correlation function~:~
%%%
$
f_1 = -1/(2 L^6)\, \langle \cO^\prime _{12}[\gamma_5] \,\,\, \cO_{21}[\gamma_5] \rangle \, .
$
%%%
%It is interesting to note that the sea sector is made out of 
%$N^{\rm sea}_{\rm f}=2$ flavours while the correlator $\cF$  in
%\req{eq:f4-corr} involves  $N^{\rm val.}_{\rm f}=3$ valence quarks.
The renormalized ratio $h_R$ can be written as follows:
%%%%
\begin{equation}
h_{\rm R}(x_0;\mu) =
\ZVApAV{}(g_0,a \mu)\,  h(x_0;g_0)\ ,
\end{equation}
%%%%
where the renormalization factor is fixed by imposing the renormalization condition~:
%%%%
\begin{equation}
\label{eq:rencond}
\ZVApAV{}(g_0,a \mu=1/L)  h(x_0=L/2;g_0)=
h_{\rm s}(x_0=L/2;g_0) \bigg \vert_{g_0 = 0}  \,\, ,
\end{equation}
%%%%
{\it i.e.} at tree level $\ZVApAV{}=1$. The renormalization condition is taken
 at the scale $\mu=1/L$ and therefore at fixed renormalized
coupling $u=\gbar^2(1/L)$. Note that as the  quarks are massless, once the continuum limit $a\to 0$ is taken, the only 
remaining scale is $L$.

The running of the scale-dependent factor $\ZVApAV{}(g_0,a \mu)$ is implemented
 in the SF formalism  via the step scaling
function (SSF), defined in the continuum as:
%%%%
\begin{equation}
\label{eq:ssfLat}
\sigVApAV{}(u) = \lim_{a \to 0} \SigVApAV{}(u,a/L) \, , \hspace*{1cm}
\SigVApAV{} (u,a/L) = 
\frac{\ZVApAV{}(g_0,a/2L)}{\ZVApAV{}(g_0,a/L)}
\Bigg \vert_{m=0,~\gbar^2(1/L) = u } \,\, .
\end{equation}
%%%%
The SSF can be written in terms of the 
evolution function $U$:~
%({\it cf.}  \req{RG_evolution}):~
%%%%
%\begin{equation}
%\label{eq:rgInt}
$
\sigVApAV{}(u) = \UVApAV{} ({1}/{2L},{1}/{L})
   \, .
$
%\end{equation}
%%%%
The SSF is
 used to run the operator $\oVApAV{O}$ between
two scales differing by a factor of two. By iterating this procedure
the running of the operator can be performed over
a large range of scales.

\section{Non-perturbative study of the step scaling function}

\begin{figure*}[!t]
\begin{center}
\includegraphics[width=.85\textwidth]{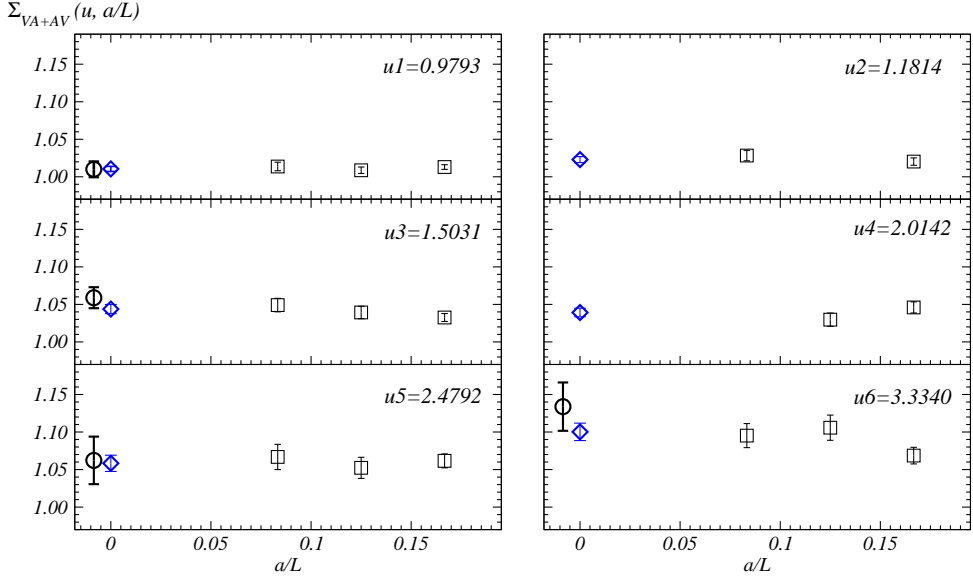}
\caption{Continuum extrapolation of the SSF $\SigVApAV{}(u,a/L)$ at fixed 
renormalized couplings $u$. The empty circles correspond to the extrapolated
 value obtained through a linear fit
 and the diamond to the one obtained via a fit
to a constant. Results are preliminary.
}
\label{fig1}
\end{center}
\end{figure*}

The computation of the SSF is performed 
with $N_{\rm f}=2$ flavours of $\cO(a)$ improved Wilson fermions. We evaluate
$\SigVApAV{}(u,a/L)$ at six values of the renormalized coupling $u$
 (labelled, in increasing order, $u1, ..., u6$). At each of these
 couplings, we consider three lattice resolutions $L/a=6,8,12$ 
 to extrapolate our data to the continuum limit. The unquenched configurations 
 employed in our computation have been previously used in the study of   
 the quark mass renormalization~\cite{DellaMorte:2005kg} (the description of our
  simulation setup can be found in this reference).
 In fig.\ref{fig1} we present the status of the ongoing
  determination of $\SigVApAV{}(u,a/L)$. 
  Data for some of the simulation points has not yet been  included as it is
 still being generated. Moreover, the statistical errors of our preliminary 
 data in fig.\ref{fig1} will  
 reduce when the complete set of configurations will be considered. The  integrated autocorrelation
 times are included in the error estimate. We  observe that the  
  autocorrelations grow when increasing the coupling and when approaching the continuum limit.
  
\begin{figure*}[!t]
\begin{center}
\includegraphics[width=.85\textwidth]{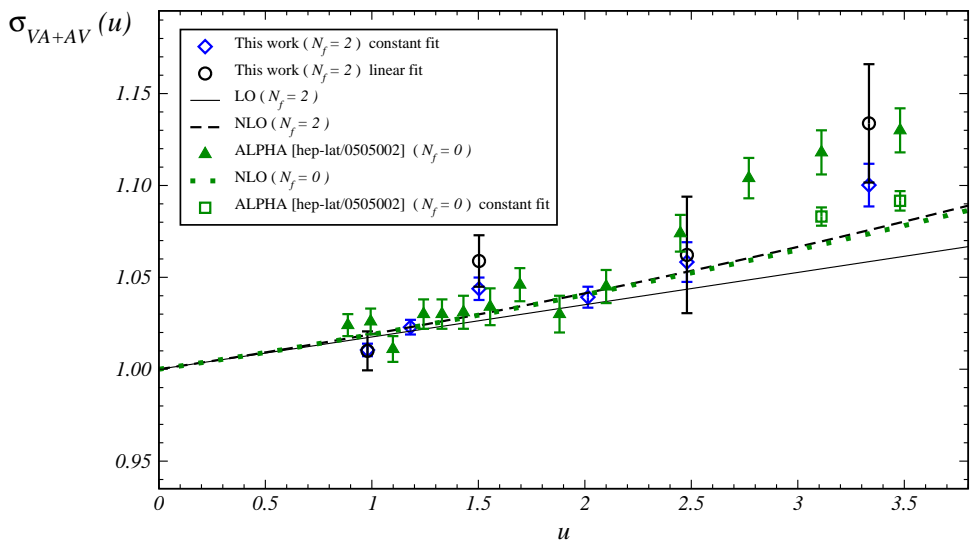}
\caption{The step scaling function $\sigVApAV{}(u)$ (discrete points) is compared 
to the LO and NLO perturbative results. In order to evaluate the quality of our data 
we compare our $N_{\rm f}=2$ results with  quenched 
data from ref.~\cite{Guagnelli:2005zc}. Results are preliminary. 
}
\label{fig_due}
\end{center}
\end{figure*}

  The $\cO(a)$ improvement of the dimension-six operator $\oVApAV{O}$ 
 has not been implemented. Although a Symanzik improvement program is possible, the mixing of
  $\oVApAV{O}$  with several 
 dimension-seven operators makes it unpractical. For each of the couplings $u$, the continuum limit
  of $\SigVApAV{}(u,a/L)$ should therefore be taken
 through a linear extrapolation. This is illustrated in fig.\ref{fig1}
  in those cases where three $L/a$ resolutions are
 already available. As our data seems to show rather small cutoff
  effects, we have also tried to fit $\SigVApAV{}(u,a/L)$  to a constant (we perform a weighted 
  average and, somehow abusively, we
  refer to it as a ``constant fit''). In
 the case of the couplings $u1, u3, u5$ and $u6$ this fit was performed by 
 discarding the $L/a=6$ data which, being far from the
 continuum, is  expected to
 contain large cutoff effects. By comparing the linear and the constant fit, we 
 observe good agreement of the extrapolated  values. A more refined analysis of the continuum 
 extrapolation will be undertaken once our complete set of data will be available.\\
%%%
%%%
In a tentative study of the quality of our data, we present in fig.~\ref{fig_due}
 the SSF $\sigVApAV{}(u)$.  The $N_{\rm f}=2$
data is compared to the quenched one from ref.~\cite{Guagnelli:2005zc}.
 The same renormalization scheme and fitting
procedures  is used in both the $N_{\rm f}=2$
 and $N_{\rm f}=0$ data (in particular, also in the quenched
  case we perform a constant fit to the continuum).~\footnote{Note that this
comparison is only intended to study the quality of our data. It is indeed improper
 to compare $N_{\rm f}=2$
 and quenched physical results at this stage since the respective 
 renormalized couplings are not taken at the same
 physical scale.} In the strong coupling regime,
  $u\sim3.5$, we observe a similar pattern in the 
 $N_{\rm f}=2$ and $N_{\rm f}=0$ data  when comparing the value of $\sigVApAV{}(u)$
 obtained through a linear and a 
 constant extrapolation to the continuum: in both $N_{\rm f}=0,2$ cases the linear extrapolation
  points lie above the constant-fit ones. Due to large 
  statistical errors, the $N_{\rm f}=2$ data shows 
 a better agreement between linear and constant fit. In fig.\ref{fig_due}, we also plot the
 perturbative expressions of $\sigVApAV{}(u)$ in both the
   $N_{\rm f}=2$ and $N_{\rm f}=0$ cases. These expressions
  were
 computed at next-to-leading order (NLO) in 
 ref.~\cite{Palombi:2005zd}. In our chosen renormalization scheme, 
 we observe a fairly good agreement between the 
 perturbative curve and the non-perturbative data 
 in the small coupling region
 and some signs of deviations in the strong coupling 
 regime. We have considered nine different  renormalization schemes 
 (for the definitions of these schemes, refer 
 to~\cite{Guagnelli:2005zc,Palombi:2005zd}). The empirical criterion 
 to identify the more appropriate schemes is to 
 consider those having a small NLO term, of the same sign as the LO one, in the
 perturbative expansion of  $\sigVApAV{}(u)$. We
  have checked that our best available scheme is indeed the one of ~\req{eq:h-corr}.
%%%
%%%
\section*{Conclusions}
We have presented the status of the computation of the non-perturbative RG running
 of the four-fermion operator $\oVApAV{O}$ using lattice QCD with two dynamical quarks. 
 This calculation will soon allow us to
 determine the universal  renormalization factor $\zrgiVApAV{;s}(\mu)$ in \req{eq:tot_renorm}. 
 The second factor on the
 r.h.s of this equation is simpler to determine, compared to $\zrgiVApAV{;s}(\mu)$,
  as it depends only on a single scale $\mu$. This 
  determination will allow to complete the renormalization 
 of the operator $\oVApAV{O}$. As our renormalization
  scheme is flavour-independent, the same renormalization 
 factors can be used to determine the $B$-parameters 
 in the strange, charm and beauty sectors.
  
\section*{Acknowledgements}
We thank Roberto Frezzotti and Francesco Knechtli for 
useful discussions. We would specially like to thank Michele Della Morte for
help and advice in the first stages of this work. 
We also thank the Computer Center of DESY-Zeuthen for their support.

\end{document}